\documentclass[lettersize,journal]{IEEEtran}
\usepackage{amsmath,amsfonts}
\usepackage{algorithmic}
\usepackage{algorithm}
\usepackage{array}
\usepackage[caption=false,font=normalsize,labelfont=sf,textfont=sf]{subfig}
\usepackage{textcomp}
\usepackage{stfloats}
\usepackage{url}
\usepackage{verbatim}
\usepackage{graphicx}
\usepackage{cite}
\usepackage{balance}
\usepackage{multirow}
\usepackage{flushend}

\usepackage{amsthm}
\usepackage{makecell}
\usepackage{threeparttable}

\begin{document}

\title{Drift-oriented Self-evolving Encrypted Traffic Application Classification for Actual Network Environment}

\author{Zihan Chen,~\IEEEmembership{Member,~IEEE}, Guang Cheng,~\IEEEmembership{Member,~IEEE}, Jinhui Li, Tian Qin, Yuyang Zhou,~\IEEEmembership{Member,~IEEE}, Xing Luan
	
\thanks{The authors are with the School of Cyber Science and Engineering, Southeast University, Nanjing 210096, China; Purple Mountain Laboratories, Nanjing 211111, China; Jiangsu Province Engineering Research Center of Security for Ubiquitous Network, Nanjing 211189, China.}}

% The paper headers
\markboth{Journal of \LaTeX\ Class Files,~Vol.~14, No.~8, August~2021}%
{Shell \MakeLowercase{\textit{et al.}}: A Sample Article Using IEEEtran.cls for IEEE Journals}

\maketitle

\begin{abstract}
Encrypted traffic classification technology is a crucial decision-making information source for network management and security protection. It has the advantages of excellent response timeliness, large-scale data bearing, and cross-time-and-space analysis. The existing research on encrypted traffic classification has gradually transitioned from the closed world to the open world, and many classifier optimization and feature engineering schemes have been proposed. However, encrypted traffic classification has yet to be effectively applied to the actual network environment. The main reason is that applications on the Internet are constantly updated, including function adjustment and version change, which brings severe feature concept drift, resulting in rapid failure of the classifier. Hence, the entire model must be retrained only past very fast time, with unacceptable labeled sample constructing and model training cost. To solve this problem, we deeply study the characteristics of Internet application updates, associate them with feature concept drift, and then propose self-evolving encrypted traffic classification. We propose a feature concept drift determination method and a drift-oriented self-evolving fine-tuning method based on the Laida criterion to adapt to all applications that are likely to be updated. In the case of no exact label samples, the classifier evolves through fully fine-tuning continuously, and the time interval between two necessary retraining is greatly extended to be applied to the actual network environment. Experiments show that our approach significantly improves the classification performance of the original classifier on the following stage dataset of the following months (9\% improvement on F1-score) without any hard-to-acquire labeled sample. Under the current experimental environment, the life of the classifier is extended to more than eight months.
\end{abstract}

\begin{IEEEkeywords}
Encrypted traffic classification, Concept drift, Self-evolving fine-tuning architecture, Windowed multi-threshold accumulation measurement
\end{IEEEkeywords}

\section{Introduction}
\IEEEPARstart{W}{ith} the rise of network security and privacy protection awareness, the encryption of network traffic has become an inevitable trend \cite{1Zeng2021Research}. In the backbone network environment with Tbps bandwidth, more than 95\% of traffic is encrypted, and some Internet services have reached almost 100\% encrypted \cite{2Chen2023Survey}. To solve the problem of encrypted traffic not being matched in plaintext to support network management and security \cite{3Xu2022Seeing}, relevant research on encrypted traffic classification has been produced. Encrypted traffic classification aims to use the remaining information disclosure not covered by encryption network protocols to classify or identify descriptive labels such as services, applications, and behaviors underneath encrypted traffic from a macro perspective without deciphering \cite{4Shen2024Machine}.\par

The existing research on encrypted traffic classification mainly focuses on deep learning frameworks and open-world environments, including classifier optimization and encrypted traffic feature engineering. Classifier optimization mainly focuses on deep learning model fusion, graph neural networks that are more suitable for expressing the interactive features of traffic, and the introduction of the latest large language models. These studies can better mine the distinguishing information in the features to better approximate the upper limit of the classification effect. Unlikely, the feature engineering of encrypted traffic enhances the initial adaptability of features from the perspective of increasing the upper limit of theoretical information gain in feature space to increase the classification performance.\par

However, compared with the closed-world environment, although unknown applications and incomplete sample coverage are considered in the open-world environment, the application update in the actual network environment should be considered. By studying the encrypted traffic at the border of the provincial backbone network, we find that the network application is constantly updated. In addition, the network environment is constantly changing, resulting in the new encrypted traffic sample features being inconsistent with those of the original samples in the same category. The changes are time-persistent, continuously decreasing the trained model's classification effect. This phenomenon is called the concept drift of encrypted traffic \cite{5Chen2023Classify}. Although encrypted traffic feature engineering can improve the initial tolerance of the method to concept drift, the concept drift is spawned very fast \cite{6Lu2019Concept}, resulting in the rapid failure of the classifier. The cost of retraining a new encrypted traffic classifier is really high due to the massive calculation and elusive labeled sample acquirement \cite{7Chen2022Length}. Repeated retraining in a short period is unacceptable in terms of computing power cost, and there is not enough time to collect enough samples. As a result, the encrypted traffic classification method has been challenging to apply in the actual network environment.\par

To solve the above problems, we propose the self-evolving encrypted traffic classification, which aims to resist the continuous failure of the model caused by the irresistible feature concept drift through the model's catch-up self-evolving. First, we study the factors that cause feature concept drift in an actual network environment, and from its universality, we propose a drift-oriented self-evolving architecture. We propose a windowed multi-threshold accumulation measurement method to solve the problem of concept drift determination in this architecture. For the self-evolving of the classifier, we introduce the Laida criterion to retrieve and label samples with high enough \textit{Softmax} confidence in the label-free prediction process, and rely on such samples (named \textbf{silver samples}) to conduct Fully Fine-Tuning (FFT), and finally significantly extend the time interval between two necessary retraining, to achieve the self-evolving encrypted traffic classification.\par

The main contributions of this paper are as follows:\par
\begin{itemize}
	\item For the first time, we pay direct attention to concept drift in encrypted traffic classification in a actual network environment, investigate its causes, and propose windowed multi-threshold accumulation measurement method to determine whether concept drift occurs in the current classifier.
	\item We propose the self-evolving encrypted traffic classification for the first time. Aiming at the different degrees of feature concept drift that may exist in all applications, we introduce the Laida criterion and propose a drift-oriented self-evolving fine-tuning method based on the high-confidence silver samples continuously recovered in the classification process, which significantly extends the effective life cycle of the model (The current dataset can prove to extend to 8 months, which may be longer.), diluting the long-term deployment costs of encrypted traffic classifications.
\end{itemize}

The structure of this paper is as follows. Section II introduces the latest research in encrypted traffic classification. In Section III, we analyze the source of concept drift in a actual network environment and introduce the Laida criterion to gather silver samples. Section IV focuses on two key links in the drift-oriented self-evolving fine-tuning of encrypted traffic. In Section V, experiments on self-evolving classification are carried out. Finally, the paper is summarized in Section VI.

\section{Related Work}
Current research in encrypted traffic classification can be divided into two parts: deep learning classifier optimization and anti-concept drift feature engineering.\par

With the development of neural networks and deep learning technology, some researchers began using deep learning methods to classify encrypted traffic. The most significant advantages of deep learning are that it does not need to rely on prior expert knowledge of feature engineering, its end-to-end learning features can directly feed raw encrypted traffic into the neural network for training and classification, and neural networks have better generalization ability than traditional machine learning methods.\par

The latest research on deep learning classifier optimization focuses on the fusion of tensor-based models (I$^2$RNN \cite{9Song2023I2RNN}) and the introduction of graph neural network models \cite{10Shen2021Accurate}. The advantage of tensor-based model fusion is that in the case of Large Language Models (LLM), with a small volume, it can play the advantages of different model architectures to construct a model with better capability, which is an essential basis for the encrypted traffic classification. To some extent, the graph neural network uses packet interactivity in the process of encrypted traffic transmission. It is superior to the tensor neural network in terms of accuracy at the cost of more calculations. However, in approaching the theoretical optimal effect, both of them also pay a considerable performance cost.\par

Although some studies began to explore the explainability \cite{8Nascita2023Improving}, there is still a problem of unexplainable features \cite{11Shen2021Fine} and the results of automatic feature selection are only sometimes preferred, mainly due to the non-high-dimensional optimization of features and the neglect of structural features.\par

On the other hand, some studies focused on anti-concept drift feature engineering to improve the upper limit of the model's initial classification accuracy. At present, the most representative studies focus on length sequence features, such as packet length sequence features \cite{12Yun2023TLS}, multi-flow length sequence features \cite{7Chen2022Length}, and packet length path signature features \cite{3Xu2022Seeing}.\par

In summary, with the support of feature engineering, the enhanced length sequence features can resist concept drift better than other features. However, these features are still static, and no matter how good the feature is, it will gradually become invalid with the continuous update of massive applications. Static classifiers cannot keep up with the feature concept drift caused by application updates and cannot be used in an actual network environment.\par

\section{Continuous Sample Acquisition under Concept Drift}
\subsection{The Negative Impact of Actual Network Environment on Existing Encrypted Traffic Classification Methods}
The concept drift of encrypted traffic features refers to the phenomenon where the representation of a feature changes arbitrarily without any change in the input (still belonging to the same feature category). It is due to the self-change or external influence on the target being classified over time. Therefore, the sources of the concept drift in encrypted traffic features are various variable factors in the actual network environment. From the perspective of encrypted traffic, it can be broadly categorized into feature concept drift caused by changes in protocol headers and protocol bodies.\par

Changes in protocol headers are mainly shown in encryption transport protocols and application layer protocols, which are covered/encrypted. On the Internet, the most typical example of the former is the TLS-1.3 protocol, while the latter is the HTTP-2.0 protocol. QUIC, as a transport layer protocol with built-in encryption capabilities, combines the two types of changes. The core impact comes from the network protocol itself.\par

Changes in protocol bodies are more diverse, including changes in the transmitted data and network environment. For example, changes in the transmitted data come directly from application updates, including data-side and functional-side updates. Differently, changes in the network environment are more extensive, including changes in hardware models, network configurations, spatial locations, and other human-selected changes, as well as changes caused by technological updates and iterations in related fields. We have summarized these two types of changes in Table \ref{tab_dl_classifiers} below.\par

\begin{table*}[!t]
	\caption{The negative impacts of protocols and the environment on the classification of encrypted traffic applications}
	\centering
	\begin{tabular}{|c|c|c|}
		\hline
		\textbf{Influencing Factors} & \textbf{Factor Name}& \textbf{Specific Negative Impacts} \\
		\hline
		\multirow{8}{*}{Protocol}&	\multirow{2}{*}{TLS-1.3}&Undetermined start of flow for an application behavior\\
		\cline{3-3}
		&	&Concept drift to TLS-1.2 statistical features\\
		\cline{2-3}
		&	\multirow{2}{*}{HTTP-2.0}& Flow-level feature confusion due to multiplexing\\
		\cline{3-3}
		&	&  Hidden application layer header length shorten continuously with access\\
		\cline{2-3}
		&	\multirow{3}{*}{QUIC}&The classifier for HTTPS traffic becomes invalid directly\\
		\cline{3-3}
		&	&  QUIC is constantly updated rapidly, so the features are constantly disturbed \\
		\cline{3-3}
		&	&  More complex multiplexing, interactivity features interference \\
		\cline{2-3}
		&Multi-protocol Application&The highly-precised single protocol classifier fails\\
		\hline
		\multirow{4}{*}{Environment}&	Rapid Updates of Applications&Features are constantly changes, new functions, new features\\
		\cline{2-3}
		&	Application Homogenization Competition& Overlapping functions of applications, increasing feature confusion\\
		\cline{2-3}
		&	User Habit Difference&  Spatially inconsistent traffic features of the same application\\
		\cline{2-3}
		&	Computational Ocean and LLM&Dramatically increasing the speed of function update and new application generate\\
		\cline{2-3}
		\hline
	\end{tabular}
	\label{tab_dl_classifiers}
\end{table*}

In summary, if we want to implement self-evolving encrypted traffic application classification, we need to consider all of the above. However, since feature concept drift is uncontrollable and difficult to predict (especially in the case of a large number of categories), the concept drift happening must be determined.\par

\subsection{Laida Criterion and Silver Samples}
When the drift representation accumulates to a certain extent, we will consider that the model is no longer effective in the current network environment and urgently needs  fine-tuning (even if some categories still have acceptable classification effects). The prerequisite for fine-tuning is that there are some new labeled samples, but the cost of collecting labeled samples through controllable end devices is very high and cannot guarantee the coverage of after-concept-drift samples. On the other hand, the traffic throughput of the analysis point is much higher than that of an end device. The single-slot edge network traffic processing device at the edge of the backbone network has a peak of about 400 Gbps throughput, and the entire device can reach up to 25.6 Tbps. On the contrary, the peak throughput of a controllable end device is only about 100 Mbps, and the average labeled sample traffic collection throughput is less than 10 Mbps. Hence, we must obtain new labeled samples directly at the analysis point.\par

However, the analysis point cannot obtain samples with completely real labels; that is, the result obtained by the classification method may not be 100\% correct due to the nature of deep learning.\par

In mathematical statistics, the Laida criterion refers to the interval calculation based on the standard deviation probability when it is assumed that a dataset is approximately normally distributed and only has random errors. For encrypted traffic classification, the \textit{Softmax} function can convert the parameter value of the neural network's penultimate layer into the confidence distribution of the category. Although the classification of a certain application can meet the sufficient measurement condition, the confidence distribution of the category may not conform to the normal distribution, depending on the nature of the classifier, especially in the case of multi-classification. The gradient saddle point on the side of the category in question may not be symmetrical.\par

Nicely, due to the arbitrary direction of concept drift, we can assume that the scope of concept drift at the next time point conforms to the normal distribution. Therefore, we can extend the assumption that the change in the \textit{Softmax} function's confidence in a certain category is consistent with the normal distribution, so we can use the Laida criterion as the verification standard. Samples with positive offset that exceed this range will be regarded as effective fine-tuning samples, which are named \textbf{silver samples} in this paper (as opposed to deterministic labeled samples in the normal construction of label datasets, which are called gold samples). Therefore, the confidence standard of the silver sample was selected as 0.997 (3 $\sigma$); that is, if the maximum value of \textit{Softmax} confidence of the current sample classification result is higher than 0.997, it is regarded as a silver sample, which will be used for the subsequent round of fine-tuning.

\section{Drift-oriented Encrypted Traffic Self-evolving Fine-tuning}
With the development of time, the features reflected by all application samples may change compared with the original sample features of this category, resulting in the inevitability of concept drift. However, simultaneously, the degree and cycle of different application updates or version iterations are not aligned, which leads to uncontrollable concept drift. Therefore, drift-oriented self-evolving is proposed in this paper, which aims to continuously fine-tune the model according to the concept drift of a particular category without considering the specific classification target.\par

\subsection{Concept Drift Determination based on Windowed Multi-threshold Accumulation Measurement}
At a macro level, the self-evolving encrypted traffic application classification plays a crucial role in maintaining the overall accuracy. This process is reflected in the fact that the accuracy of each category needs to be improved, and the more the category with concept drift, the more it needs to be improved.\par

Due to the arbitrariness of concept drift in all directions and the continuity of application updates, the classifier intuitively feels that the current sample features are "similar enough but not very similar" to the features during training, which is reflected in the decline of classification confidence. It is worth noting that for classifiers, a decline in classification confidence is a necessary condition for concept drift to have an impact. Therefore, judging concept drift only by the appearance of classification confidence decline is wrong.\par

In order to make better use of classification confidence decline to judge concept drift, we propose a concept drift determination method based on windowed multi-threshold accumulation measurement to determine whether a specific category or the whole classifier have concept drift (whether it needs to be fine-tuned).\par

The purpose of determining whether concept drift occurs in a certain category is that the update cycle of some applications is concise, and the degree of concept drift per unit time is much higher than that of other categories. Although it has little impact on the model's overall accuracy, it is invalid for this category. Hence, the overall classifier's determination is that most categories are invalid to a certain extent, so the results obtained by the model do not have a good reference value.\par

\begin{figure*}[!t]
	\centering
	\includegraphics[width=\linewidth]{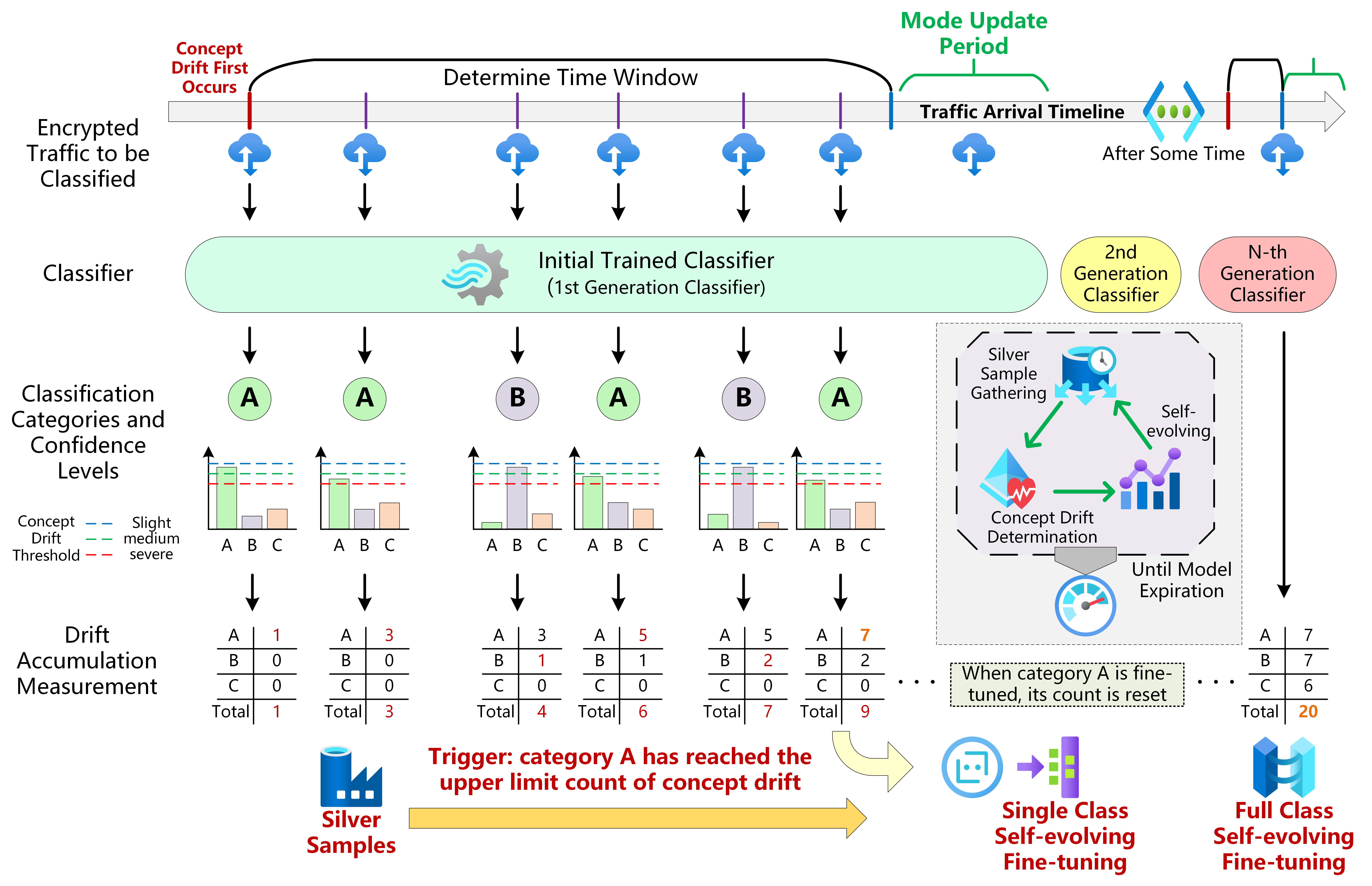}
	\caption{Conceptual drift determination process diagram based on windowed multi-threshold accumulation measurement}
	\label{fig_wmam}
\end{figure*}

Therefore, according to the theory in Section III, we set multiple thresholds for each category and the overall model, respectively, and there are different scores for reaching different thresholds. By calculating the accumulative score of the new classified sample in the current time window, we can determine whether the category or the whole model needs to be fine-tuned, as shown in Figure \ref{fig_wmam}.\par

In particular, since the concept drift determination should not affect the standard classification of encrypted traffic, it is impossible to constantly slide to calculate whether concept drift occurs in a specific time slice (otherwise, it will bring huge polling performance overhead). Since the overall accuracy decline of the model comes from the accumulation of class accuracy decline, we will carry out the number of accumulative times when the confidence is too low for the samples of a certain class (assumed as class \textbf{\textit{A}}) for the first time, and the lower the threshold, the more times will be accumulated. If the accumulative number of measurements in the current time window reaches the judgment threshold, it is regarded as the happening of concept drift. A more severe concept drift is considered to have occurred if a lower decision threshold is reached. Subsequent fine-tuning can be done depending on the degree of conceptual drift in the current category.\par

Suppose the single class of the current model has only slight concept drift at worst. Still, many classes already have concept drift (usually in the model after long-term use, which is also determined by accumulation measurement). In that case, the entire model must be fine-tuned to adapt to the current feature representation distribution.\par

\subsection{Drift-oriented Self-evolving Fine-tuning}

Compared with the task specialization of an LLM, the initial model in the self-evolving lifecycle can be regarded as a pre-training process and the subsequent evolution process as a fine-tuning process of the model. However, the target dataset is derived from the silver samples continuously obtained in the classification process, and fine-tuning is constantly occurring.

Therefore, in view of this paper's self-evolving scenario, we propose drift-oriented self-evolving fine-tuning. It is worth noting that "drift-oriented" means that we will not specifically study the feature changes of a certain category during the whole lifecycle of the model but let the classifier directly be fine-tuned when there is a concept drift. This adaptability to concept drift is also the meaning of 'self' in self-evolving, and it ensures the model's accuracy over time.\par

It's crucial to understand that, in the context of encrypted traffic classification, self-evolving is inherently model-independent. This means that any deep learning model capable of fine-tuning can be seamlessly integrated into the self-evolving fine-tuning architecture, providing a high level of flexibility and adaptability. Unlike sequence classification in NLP, where Transformer are typically used, encrypted traffic classification requires models that can maintain high throughput network requirements. Therefore, for the classification model to be fine-tuned, we have chosen LS-LSTM \cite{7Chen2022Length} from previous research to demonstrate the enhancement potential of our proposed method.\par

Since the target model does not necessarily have the ability to adapt to prefixes, we chose FFT over other methods. 

\section{Experiment Evaluation}

\subsection{Dataset and Experimental Environment}

Since the current public dataset cannot effectively support the research of self-evolving encrypted traffic classification, we conducted traffic sample collection in the actual network environment of Jiangsu Province (in order to ensure privacy, the specific information about traffic collection personnel, device model, and account is not disclosed) from September 27, 2023, to July 16, 2024. It includes the traffic of various Internet applications and web pages, mainly multimedia-related traffic, a total of 372 GB (the content are relatively stable and can better reflect the concept drift of traffic features). The silver sample dataset still uses 80\% of the samples as training and 20\% as test samples as primitive datasets.\par

\begin{table*}[!t]
	\caption{Statistics of the datasets (present by flow count)}
	\centering
	\begin{threeparttable}
		\begin{tabular}{|c|c|c|c|c|c|c|}
			\hline
			\textbf{Application} & \makecell{\textbf{Initial training}\\ \textbf{samples}}& \makecell{\textbf{1st stage} \\ \textbf{total samples}}& \makecell{\textbf{1st fine-tune}\\\textbf{silver samples}}\tnote{*} & \makecell{\textbf{2nd stage}\\\textbf{total samples}}& \makecell{\textbf{2nd fine-tune}\\\textbf{silver samples}}\tnote{*}& \makecell{\textbf{Final stage}\\\textbf{long-span samples}}\\
			\hline
			Bilibili&4585&4587 &1407 &4587 &2340 &1529 \\
			\hline
			Douyin&	125&123&21 &123 &34 &41 \\
			\hline
			MGTV&1965& 1963 &159 &1963 &371 &656 \\
			\hline
			Youku&4425	&4427  & 825&4427 &1096 &1477 \\
			\hline
			QQ Music&47828&47830  &39099 &47830 &40542 &15944 \\
			\hline
			IQiYi&2925&  2923  &1809 &2923 &2304 &977 \\
			\hline
			\textbf{Total}&61853	&61853&43320 &61853&46687 & 20624\\
			\hline	
		\end{tabular}
		\begin{tablenotes}
			\footnotesize
			\item[*] The silver samples are gathered by the previous one version classifier, which is model-dataset specific.
		\end{tablenotes}
	\end{threeparttable}
	\label{tab_dataset}
\end{table*}

It is worth noting that to reflect the changes in application characteristics, we deliberately selected six applications with large user and traffic volumes. As shown in Table \ref{tab_dataset}, silver samples were calculated only for the first and second fine-tuning stages (the number is determined by the actual experimental results). In contrast, the long-span third classification stage's samples had a more extended period than the previous two. In order to reflect the continuity of the traffic label sample, we did not divide the data strictly according to the date. However, we carried out proportional segmentation according to the continuity of the time axis according to the sample size. At the same time, we directly used datasets with a highly uneven number distribution to reflect the differences in the sample size distribution of different applications in actual network environments. Datasets will be published at \textbf{https://data.iptas.edu.cn/web/tbps} after a strict privacy audit.\par

Since the determination of concept drift is a simple statistical process of sample classification confidence results, the preliminary experiment found that the threshold is related to the throughput of current network traffic (that is, to the total number of samples to be classified). Therefore, we directly assume that each stage triggers the determination of concept drift in the following experiments. Therefore, each classification stage must be self-evolving as it transitions to the next stage.

\subsection{Encrypted Traffic Classification Effect Diminishing and Self-evolving Classification Experiment}

At the same time, we conducted a diminishing experiment and a self-evolving experiment on the classification effect of encrypted traffic. In this experiment, we first compared the effects of the initial trained model, the fine-tuned model from initial trained model using the silver samples from the first stage dataset (Level-1 fine-tuned), and the fine-tuned model from Level-1 fine-tuned model using the silver samples from the second stage dataset (Level-2 fine-tuned) on the data subset in the next stage, respectively. Then we fine-tuned the model in the current stage and tested the results after fine-tuning. The \textit{epoch} of initial training or fine-tuning was 50 rounds, the \textit{batch-size} was 500, and the \textit{learning-rate} was fixed at 0.0025. We conducted ten rounds of experiments, and each round of the training set and test set was randomly divided in equal proportion (80\% and 20\%). The average results are shown in Table \ref{tab_experiment}.\par

\begin{table*}[hbpt]
	\caption{The performance of the three different stage models under their own stage datasets, next-stage datasets, and final-stage datasets}
	\centering
	\begin{tabular}{|c|c|c|c|}
		\hline
		 & \textbf{Initial trained LS-LSTM}& \textbf{Level-1 fine-tuned LS-LSTM} &\textbf{Level-2 fine-tuned LS-LSTM}\\
		\hline
		Training F1-score&	0.9321&0.9902 &0.9960 \\
		\hline
		Testing F1-score& 0.9189& 0.9619& 0.9822\\
		\hline
		F1-score of Next-stage Samples&	0.8960& 0.8956 & N/A\\
		\hline
		F1-score of Final Samples&0.5950&0.5842 &0.5898 \\
		\hline
		Silver Sample Percentage in Final Samples& 41.66\%& 56.74\%&68.20\%\\
		\hline
	\end{tabular}
	\label{tab_experiment}
\end{table*}

The following conclusions can be drawn from the experimental results:\par
\begin{enumerate}
	\item By comparing the classification performance of each round of initial training or fine-tuned model in the current data subset and the following data subset, concept drift exists, which is not only reflected in the decline of the classification confidence of some samples on the micro level but also in the decline of the F1-score of the overall dataset on the macro level. Therefore, the diminishing of the encrypted traffic classification effect over time is natural.\par
	
	\item Compared with the initial training model and the two fine-tuned models, the catastrophic forgetting that often occurs in the field of NLP does not appear. However, it can be seen that the training F1-score is constantly improving, while the classification F1-score of the next round of data is constantly decreasing. At the same time, although the silver sample rate of the final test dataset is increasing (that is, the accuracy of the classification of the sample is high confidence or the classification of the sample is firmly wrong), the F1-score even shows a decline in general. This indicates some overfitting, which is expected because self-evolving fine-tuning cannot give the model an infinite lifetime. It can only extend the lifetime of the model. With second stage experimental data as a calculation, the life cycle is extended to 8 months (compared to a first-order model with a span of only 2 months).\par
	
	\item Comparing the F1-score of the next round of data classification before fine-tuning with the test F1-score of the model after fine-tuning, it can be found that self-evolving fine-tuning is effective and can significantly improve the model's ability to adapt to new data. It is worth noting that because the samples of each stage have a specific span, the fine-tuning sample set actually covers the data of the current time node (that is, the update of the application are continuous). Therefore, using the silver sample as the fine-tuning dataset for self-evolving is feasible and reasonable.\par
	
	\item For the long-span third classification samples tested in the final test, the initial and fine-tuned models performed poorly and showed a downward trend on the whole. The main reason is that the feature concept drift of the long period (up to 3 months with the second stage dataset) could not be expressed by the current fine-tuning, especially for some categories with few samples. The weakness of feature expression activation makes it more challenging to adapt to changes, so self-evolving needs to be constantly deployed. On the other hand, the average F1-score of the Level-2 fine-tuned model is slightly increased. This may be because the second stage data is closer to the final data in time (although still far away), and the features are slightly more similar, proving the continuity of application updates.\par
\end{enumerate}

\section{Conclusion}
This paper proposes a self-evolving encrypted traffic application classification method for the actual large-scale Internet environment, which does not need to consider the specific changes of the actual classification objectives, nor does it need to restrict the role of the model. It directly relies on the silver samples accumulated in the prediction process, based on the concept drift determination through windowed multi-threshold accumulation measurement, and constantly fine-tuning the model. Thus, the life cycle of the classification model can be extended greatly without any real labeled samples and retraining.\par

In the subsequent research, the relationship between concept drift determination and sample size (network traffic throughput) will be further refined first, and the relationship between the \textit{Softmax} threshold value of silver sample determination (currently 0.997 directly) and model/sample needs to be further studied. In addition, the dataset will be extended to quantify the extension of the lifecycle and the cost that can be really reduced.\par

\section*{Acknowledgments}
This paper is supported by the Youth Fund of the National Natural Science Foundation of China under Grant Number 62402101, the Joint Funds of the National Natural Science Foundation of China under Grant Number U22B2025, the Jiangsu Funding Program for Excellent Postdoctoral Talent under Grant Number 2024ZB494. This paper is part on the topic of the encrypted traffic classification. Guang Cheng is the corresponding author.

\bibliographystyle{IEEEtran}
\bibliography{IEEEabrv,self_evo_bib}

\vskip-0.52in

\begin{IEEEbiographynophoto}{Zihan Chen}
	obtained his Ph.D. degree in Cyber Security from Southeast University in 2023 and B.S. degree in Software Engineering from Central South University in 2017. He is currently working as a postdoc with the School of Cyber Science and Engineering at Southeast University. His major research interests include cyber security, encrypted traffic classification, encrypted traffic feature engineering, and deep learning. He is a Member of IEEE and works as a reviewer for multiple Journals such as IEEE IoTJ and the duty editor of the Journal of Cyberspace.
\end{IEEEbiographynophoto}

\vskip-0.2in

\begin{IEEEbiographynophoto}{Guang Cheng}
	received his B.S. degree in Traffic Engineering from Southeast University in 1994, his M.S. degree in Computer Application from Hefei University of Technology in 2000, and his Ph.D. degree in Computer Network from Southeast University in 2003. He is a Full Professor in the School of Cyber Science and Engineering, Southeast University, Nanjing, China. He has authored or coauthored seven monographs and more than 100 technical papers, including top journals and top conferences. His research interests include network security, network measurement, and traffic behavior analysis. He is a Member of IEEE and a Senior Member of CCF.
\end{IEEEbiographynophoto}

\vskip-0.2in

\begin{IEEEbiographynophoto}{Jinhui Li}
	received his B.S. degree in Cyber Security from Southeast University in 2021. He is currently a Master's student at the School of Cyber Science and Engineering, Southeast University. His major research interests include feature engineering and classification of multimedia traffic, as well as encrypted traffic classification.
\end{IEEEbiographynophoto}

\vskip-0.2in

\begin{IEEEbiographynophoto}{Tian Qin}
	received the B.S. degree in information	and computing sciences from HoHai University (HHU) in 2020. He is currently a Doctor candidate at the School of cyber science and engineering,
	Southeast University, Nanjing, China. His current research interests include network traffic detection and federated learning.
\end{IEEEbiographynophoto}

\vskip-0.2in

\begin{IEEEbiographynophoto}{Yuyang Zhou}
	is currently working as a postdoc with the School of Cyber Science and Engineering, Southeast University. His major research interests include moving target defense, Android malware detection, and security modeling.
\end{IEEEbiographynophoto}

\vskip-0.2in

\begin{IEEEbiographynophoto}{Xing Luan}
	received the B.S. degree in computer science and technology from HoHai University (HHU) in 2023. He is currently a Doctor candidate at the School of cyber science and engineering,
	Southeast University, Nanjing, China. His current research interests include network traffic detection and VPN traffic.
\end{IEEEbiographynophoto}

\vspace{11pt}

\vfill

\vfill

\end{document}